# ESTIMATING $\beta \equiv \Omega_\circ^{0.6}/b$ FROM THE LOCAL GROUP & CLUSTER PECULIAR VELOCITIES

Manolis PLIONIS

*International School for Advanced Studies, Via Beirut 2–4, 34014 Trieste, Italy & National Observatory of Athens, Lofos Nimfon, Thesio, 11810 Athens, Greece*

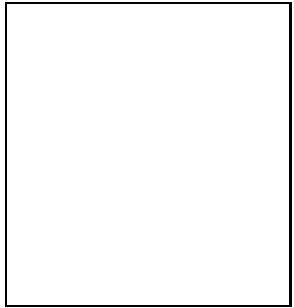

**Abstract**

Comparing the observed peculiar velocities of the Local Group and of nearby Abell clusters of galaxies with those predicted using linear perturbation theory and the 3-D reconstruction algorithm of Branchini & Plionis (this volume) the value of $\beta$ can be estimated. I show that the anisotropy that causes the LG motion does not only extend to very large depths ($\sim 160\ h^{-1}$ Mpc) but it is also very coherent, which suggests that the value of $\beta \approx 0.21 \pm 0.015$, obtained from the LG velocity, should not be significantly affected by cosmic variance. Comparing the Abell cluster and QDOT galaxy dipoles in real space their relative biasing factor are estimated to be $b_{cI} \approx 3.5 \pm 0.5$. Finally, using a suitable set of available cluster peculiar velocities I obtain a somewhat lower value of $\beta \approx 0.15 - 0.17$ but with a large uncertainty ($\gtrsim 0.1$) when weighted by the large observational errors. More and better cluster peculiar velocity data are needed to reduce this uncertainty.

## 1 Introduction

Under the linearity assumption (of gravity and biasing) linear perturbation theory can be used to relate the peculiar velocities of cosmic objects (ex. galaxies and clusters) with their peculiar accelerations, usually estimated by their dipole moments, to determine the cosmological $\beta$-parameter [$\beta \equiv \Omega_\circ^{0.6}/b$; where $b \equiv \frac{\delta N/N}{\delta M/M}$ is the biasing factor]. This relation is:

$$\beta = 4\pi \langle n \rangle \vec{v}/\vec{D} \qquad (1)$$

The estimate of $\vec{v}$ is provided in the case of the Local Group by the Doppler-interpreted dipole of the CMB radiation temperature or in the case of other galaxies and clusters by using redshift independent distance estimates (cf. [1]). The $\vec{D}$ estimate is provided by summing moments of some mass tracer distribution surrounding the source object, taking into account however their selection functions and compensating for possible sky incompleteness. Even so one should be cautious in deducing the value of $\beta$ from eq.(1) for a number of other reasons:

- Sparse sampling introduces shot-noise that enhances $\vec{D}$, especially at large depths.

- Using one source point (the LG) entails large uncertainties due to cosmic variance.

A detailed analysis, using simulations, showed that the $\beta$-parameter estimated on the LG by using high peaks of the density field (clusters) as mass-tracers is very uncertain [6].

On the other hand, the use of many source objects could greatly reduce the latter uncertainty but unfortunately such analyses are usually hampered by the large uncertainty in $\vec{v}$ (due to large uncertainties in their estimated distances) as well as in $\vec{D}$ (due to their surrounding non-symmetric survey volume). See however [2] and [1] for such attempts.

## 2 $\beta$ from the LG $\vec{v}$ using the Abell-cluster and QDOT dipoles

Since the most accurately determined value of $\vec{v}$ is that of the LG the above test has been extensively applied to the LG by using many different mass tracer catalogues (see references in [1], [4]). The general result of these studies is that the dipole moment of all these objects is well aligned with that of the CMB, indicating that indeed they trace the underline mass distribution. However, the scale where the dipole apparently converges to its final value differs from catalogue to catalogue with an indication of a strong dependance of the convergence depth, $R_{conv}$, to the characteristic depth of galaxy catalogue, $r^*$, with $R_{conv}(r) \approx cr^*$ with $c \approx 1$, which implies a spurious convergence. Only in the cluster dipole case, where $R_{conv} \approx 160\ h^{-1}$ Mpc, we have $c < 1$, probably indicating the true final dipole convergence. The different $R_{conv}$ values could be, among other things, due to: **(a)** redshift-space distortions ($r.s.d.$) coupled with shot-noise effects, **(b)** a distance dependent biasing of the different mass tracers, **(c)** a coherent and large-scale anisotropy generating the LG motion, only parts of which are sampled by the different galaxy dipoles.

The first possibility cannot be the main explanation since in most studies redshift distortions are accounted for (cf. [5], [1], Branchini & Plionis, this volume; hereafter BP). Furthermore, in the cluster case the shot noise effects are minimal since they are volume limited much beyond $R_{conv}$. The second case lacks, at present, a physical justification while the last case has been shown to be true in [4], where it was found that the differential cluster and QDOT dipoles in large equal volume shells up to $R_{conv}$ point in the general CMB dipole direction, which implies the 'coherence'. In table 1 I present the cluster and QDOT differential dipole directions having included new cluster redshifts and after correcting for redshift-space distortions.

| Shell ($h^{-1}$ Mpc) | tracer | # | l | b | $\delta\theta_{cmb}$ |
|---|---|---|---|---|---|
| 0 - 99 | cluster/QDOT | 47/1181 | 279°/238° | 27°/26° | **4°/34°** |
| 99 - 125 | cluster/QDOT | 45/235 | 311°/96° | 5°/29° | **41°/121°** |
| 125 - 143 | cluster/QDOT | 47/106 | 299°/292° | 17°/-10° | **25°/42°** |
| 124 - 157 | cluster/QDOT | 50/65 | 266°/259° | 22°/32° | **13°/15°** |

Table 1: Differential dipole direction after correcting for $r.s.d.$ and including in the cluster dipole ($1^{st}$ shell) the effect of the Virgo cluster (see text).

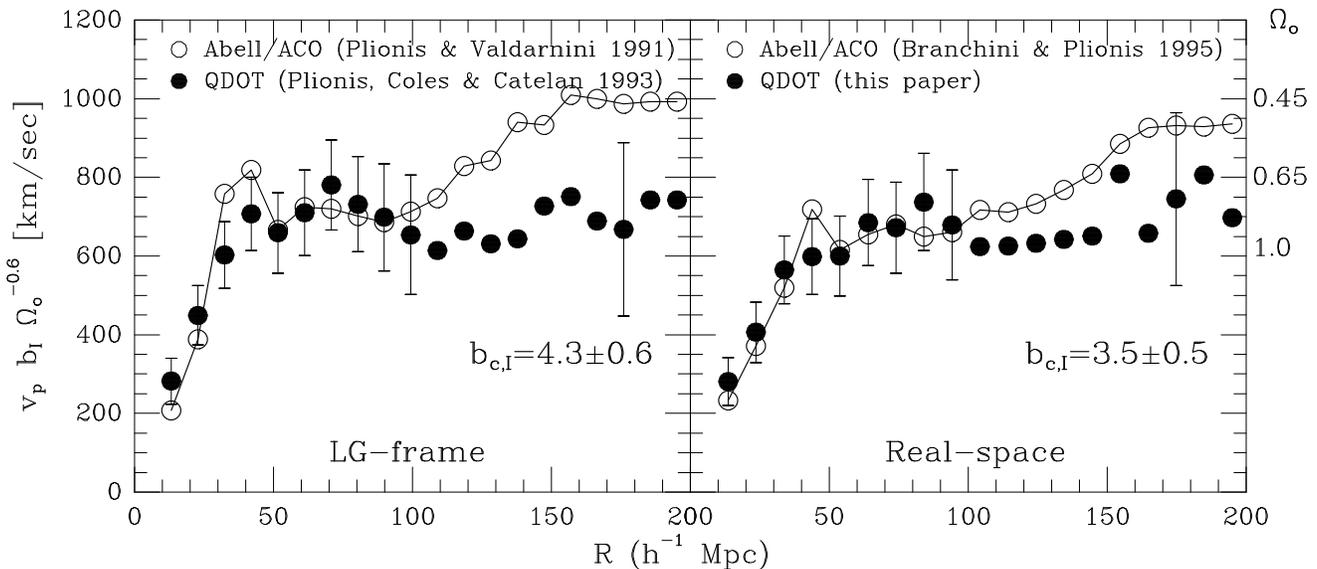

Figure 1: QDOT-galaxy and Abell-cluster dipoles, the latter rescaled by a constant *biasing* factor to *best-fit* each other within the $[10, 100]$ $h^{-1}$ Mpc region. **Left panel**: distances in LG frame. **Right panel**: *r.s.d.*-corrected distances. Errorbars are QDOT shot-noise errors (for clarity only few are plotted).

Therefore the estimated value of $\beta$ obtained from the dipoles of existing galaxy catalogues should be considered strictly an upper limit to the true value. BP using the *r.s.d.*-corrected cluster dipole find $\beta_c \approx 0.21$ with a sampling and cosmic variance uncertainties of $\sim 0.015$ and $\sim 0.05$ respectively. However, my suspicion is that due to the strong coherence of the anisotropy causing the LG motion the above value of $\beta_c$ should reflect its true value (within the sampling uncertainty).

It is interesting to compare the QDOT-galaxy and Abell-cluster dipoles within the depths where both have good sampling ($\lesssim 100$ $h^{-1}$ Mpc due to QDOT selection function) while extending the cluster dipole to shallower depths to include the effects of the Abell-like Virgo cluster (such that the virgocentric infall velocity is $\sim 200$ km/sec). Thus the lower limit of integration is set to $r \sim 10$ $h^{-1}$ Mpc. Since both trace the underline mass density field (relatively good alignment of their dipoles with that of the CMB) the only difference between their estimated dipoles should be a constant factor, reflecting their relative biasing parameter, $b_{cI}$. I plot in figure 1 the QDOT dipole (after subtracting the corresponding shot-noise one) and the rescaled, by the factor shown, Abell-cluster dipole. The agreement of the two dipoles within 100 $h^{-1}$ Mpc is excellent implying that:

**(1)** Indeed IRAS galaxies and Abell clusters trace both the underline mass distribution.
**(2)** Linear biasing is valid at least on large ($\gtrsim 10$ $h^{-1}$ Mpc) scales.
**(3)** The cluster - IRAS galaxy relative bias is $b_{cI} \approx 3.5$, in agreement with independent studies.

Furthermore, it is evident that although the cluster dipole continues growing for $r > 100$ $h^{-1}$ Mpc the QDOT one flattens out with only weak evidence for a deeper contribution (see also $4^{th}$ shell in table 1). It has been verified, using simulations [3], that this is due to the QDOT selection function which is such that the QDOT dipole would miss up to a $\sim 20\%$ total contribution if such did exist from depths $\sim 150$ $h^{-1}$ Mpc. The final value of $\Omega_\circ$ from this analysis depends on the unknown biasing factor of the IRAS-galaxy or Abell-cluster distributions with respect to that of the mass. However, with relative biassing of $b_{cI} \approx 3.5$ one would need a value of $b_I \approx 1.3 - 1.4$ to obtain $\Omega_\circ = 1$.

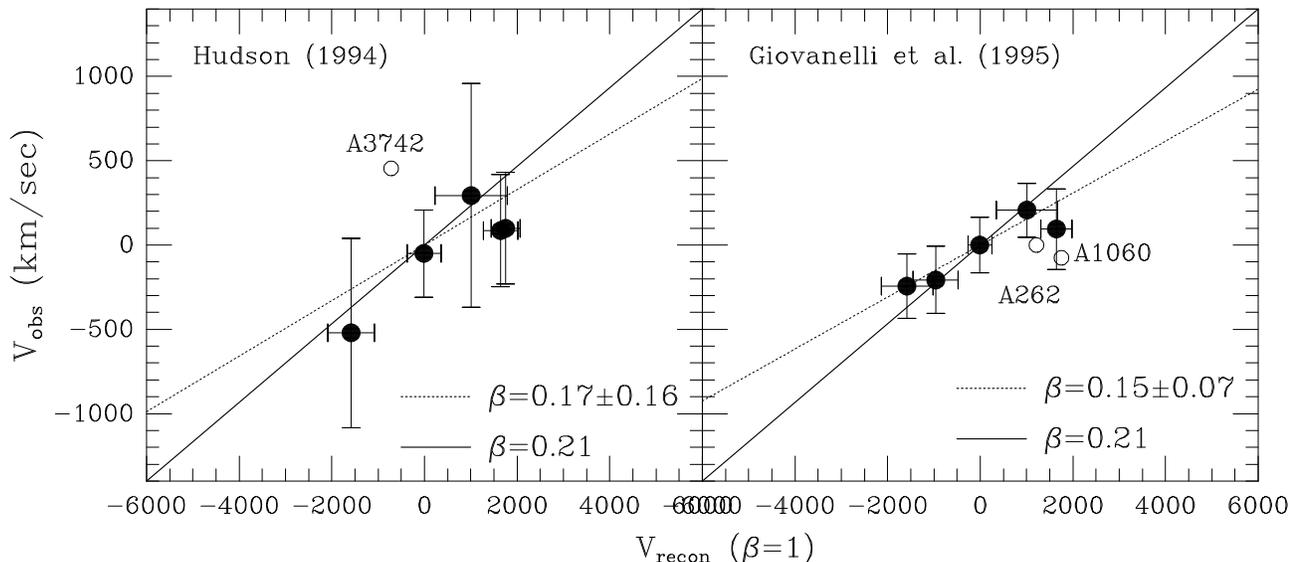
Figure 2: Observed versus predicted cluster peculiar velocities.

## 3 $\beta$ from cluster peculiar velocities

Using the reconstruction algorithm described in BP we can relate, using eq.(1), the *r.s.d.*-corrected value of $\vec{D}$ with the observed value of $\vec{v}$ for those clusters for which the latter is available (by means of $D_n - \sigma$ or *T-F* relations) and within the limit of our reconstructed volume $\lesssim 200\ h^{-1}$ Mpc. Due the the large uncertainties especially in $\vec{v}$ a large number of clusters should be used (which presently is not available). In this preliminary analysis we have chosen to exclude clusters that have uncertain peculiar velocities, either due to very discrepant $D_n - \sigma$ and *T-F* distances (ie., A262, A3742) or due to known non-linear effects (ie., Cen30).

In figure 1 the observed $\vec{v}$ is compared with that predicted by eq.(1) for clusters with *T-F* distances taken from the references indicated in the figure. There is a good correlation and the slope gives $\beta \approx 0.15 - 0.17$ which is near although lower than the value found from the LG dipole analysis (see BP). However, the number of available clusters is still too small to put strong constraints on the value of $\beta$. Note that although most of the clusters in figure 3 fall on the $\beta = 0.21$ line, the fit is biased to a lower value by Hydra (A1060) (which has 4 Abell neighbours within 30 Mpc) and A1367. If this lower $\beta$-value is verified then an $\Omega_\circ < 1$ model would be probably preferred since the IRAS galaxy biasing value that it implies is $b_I > 1.7$ (see previous section).